\documentclass[12pt, draftcls, onecolumn]{IEEEtran}
\usepackage{setspace}
\usepackage{amssymb}
\usepackage{cite}
\usepackage{mathrsfs}
\usepackage[dvips]{graphicx} 
\usepackage[cmex10]{amsmath}
\usepackage{array}
\usepackage{mdwmath}
\usepackage{mdwtab}
\usepackage[tight,footnotesize]{subfigure}
\usepackage[caption=false,font=footnotesize]{subfig}

\begin{document}

\title{Practical Analysis of Codebook Design and Frequency Offset Estimation for Virtual-MIMO Systems}

\author{\begin{large}{Jing Jiang$^\dag$,  John S. Thompson$^\S$, Hongjian Sun$^\ddag$, Peter M. Grant$^\S$}        \end{large}\\
\vspace{0.5em}
\begin{small}
$^\dag$ Center for Communication Systems Research, University of Surrey, Guildford, GU2 7XH, UK. \\
$^\S$ Institute for Digital Communications, School of Engineering, University of Edinburgh, Edinburgh, EH9 3JL, UK.\\
$^\ddag$ Department of Electronic Engineering, King's College London, London, WC2R 2LS, UK. \\ Email: jing.jiang@surrey.ac.uk; John.Thompson@ed.ac.uk; hongjian.sun@kcl.ac.uk; Peter.Grant@ed.ac.uk.\\
\end{small}
}

\maketitle

\vspace{-3.5em}
\begin{abstract}

A virtual multiple-input multiple-output (MIMO) wireless system using the receiver-side cooperation with the compress-and-forward (CF) protocol, is an alternative to a point-to-point MIMO system, when a single receiver is not equipped with multiple antennas.
It is evident that the practicality of CF cooperation will be greatly enhanced if an efficient source coding technique can be used at the relay. It is even more desirable that CF cooperation should not be unduly sensitive to carrier frequency offsets (CFOs). This paper presents a practical study of these two issues. Firstly, codebook designs of the Voronoi vector quantization (VQ) and the tree-structure vector quantization (TSVQ) to enable CF cooperation at the relay are described. A comparison in terms of the codebook design and encoding complexity is analyzed. It is shown that the TSVQ is much simpler to design and operate, and can achieve a favorable performance-complexity tradeoff.
Furthermore, this paper demonstrates that CFO can lead to significant performance degradation for the virtual MIMO system. To overcome this, it is proposed to maintain clock synchronization and jointly estimate the CFO between the relay and the destination. This approach is shown to provide a significant performance improvement.

\end{abstract}

\vspace{-0.5em}
\begin{IEEEkeywords}
\vspace{-1.2em}
Virtual MIMO system, CF cooperation, source coding technique, codebook design, effects of carrier frequency offsets.
\end{IEEEkeywords}

\IEEEpeerreviewmaketitle

\section{Introduction}

Multiple-input multiple-output (MIMO) systems have recently emerged as one of the most significant wireless techniques, as they can greatly improve spectral efficiency, channel capacity and link reliability of wireless communications\cite{Gesbert}. These benefits have encouraged extensive research on a virtual MIMO system where the transmitter has multiple antennas and each of the receivers has a single antenna\cite{Ng}\cite{Jiang_TVT1}. When the transmitter does not have perfect channel state information (CSI) for the wireless link to each receiver, which is a common scenario in practical situations, single-antenna receivers can work together to form a virtual antenna array and reap some performance benefits of MIMO systems\cite{Liang}\cite{Ding2010}.
The idea of receiver-side local cooperation is attractive for wireless networks since a wireless receiver may not be able to use multiple antennas due to size and cost limitations.
For example, suppose a customer carries two mobile terminals that are a single-antenna 3G (or 4G) enabled user device and a simple relay device.
Since the distance between the two devices is general much shorter than that from the base station,
the two closely located devices could perform cooperation through short-range Wi-Fi, Bluetooth, or Ultra-Wideband communications.
With such cooperation, the customer could expect traditional $2\times2$ MIMO benefits as if the two antennas belonged to the intended single-antenna user device.

Motivated by the above practical scenario, we consider such a cooperative virtual-MIMO system here, with one remote multi-antenna transmitter sending information to several closely spaced single-antenna receivers. Many of the techniques developed for MIMO systems can be extended to be used in this virtual-MIMO system. For example, we also implement the bit-interleaved coded modulation (BICM) technique\cite{Caire}, which introduces a spatial and temporal bit interleaver into the transmitter, to provide forward error correction (FEC) and improve system performance. But unlike point-to-point MIMO systems, we need to perform cooperation among the receivers.
As for the cooperation protocol, since the relays (i.e. the assisting receivers) are located close by the destination receiver in our scenario, compared to amplify-and-forward (AF)\cite{Krikidis2008} and decode-and-forward (DF), the compress-and-forward (CF) relay protocol provides superior performance\cite{Host-Madsen1} and therefore serves as the best candidate for this system.
Most previous work on CF cooperation has focused on the classical three terminal relay channel, such as \cite{Host-Madsen1, Qi2009,Jiang}. An extension to a virtual-MIMO system is introduced in \cite{Ng}, and the theoretical achievable cooperative capacity is analysed.
Our recent work \cite{Jiang_TVT1} and \cite{Jiang2} considers a virtual-MIMO system with CF cooperation. But reference \cite{Jiang2}  focuses
on the system performance assessment in terms of the system throughput and error probabilities. Paper \cite{Jiang_TVT1} designs an adaptive modulation and cooperation scheme when the minimum mean square error (MMSE) detection is used at  the destination. In both \cite{Jiang_TVT1} and \cite{Jiang2}, only an optimal vector quantization (VQ), i.e. the Voronoi VQ, is considered. To reduce the complexity and enhance the practicality of CF cooperation, this paper will propose an alternative VQ, the tree-structure VQ.

For conventional MIMO systems, all the antennas at the transmitter are fed with the same clock, and the same holds for the receiver side.
A major distinction of the virtual-MIMO system compared to the MIMO case is that cooperating antennas are running with different oscillator frequencies. Carrier frequency offset (CFO), which is caused by oscillator mismatch between the transmitter and the receiver, can be estimated and then compensated at the receiver in MIMO systems. The MIMO system performance will therefore not be severely degraded by CFO. However, for the virtual-MIMO system, receiver-side cooperating antennas need to estimate and compensate their CFOs independently. Their different residual CFOs will result in inter-block interference and distort the correlation properties of the signals received at the destination, so that the system performance maybe impaired significantly. Estimation of CFO in a MIMO system has been investigated in \cite{McKeown,Deng,Yao} with using different estimation algorithms. When orthogonal frequency division multiplexing (OFDM) is considered, \cite{Weng2007}  has addressed the effect of CFO on channel estimation performance, and pilot sequences have been designed in \cite{Zeng2007} and \cite{Chen2008} for joint CFO and channel estimation. The estimation methods for CFO in \cite{McKeown,Deng,Yao,Weng2007,Zeng2007,Chen2008}  are only for MIMO systems. In cooperative systems, the use of equalization has been proposed to mitigate effects from CFOs, such as in \cite{Yilmaz} and \cite{Yadav},
but only for a single-antenna transmitter. Our paper focuses on the effects of CFO in the virtual-MIMO system. To the best of our knowledge, it is the first study of the CFO effects in the cooperative system with a multi-antenna transmitter.

The main contributions of this paper are twofold. Firstly, we present a practical virtual-MIMO system that implements CF cooperation with a standard source coding technique at the relay.
To perform source coding, we consider two codebook design algorithms, Voronoi VQ and tree-structure vector quantization (TSVQ). To the best of our knowledge, it is the first time that TSVQ is applied to digital modulation signals. Their codebook design complexities and encoding complexities are investigated. Simulation results show that the TSVQ approach we designed is much simpler for encoding and more computationally efficient than the baseline Voronoi VQ. Moreover, for practical considerations, this paper studies the effects of CFO, and demonstrates that CFO can lead to severe performance degradation for the virtual MIMO system. To overcome these effects, a clock synchronization and joint CFO estimation scheme is proposed, to exploit the benefits of MIMO CFO estimation. Simulation results show that the proposed scheme provides a significant performance advantage.

The paper is organized as follows: Section \ref{System Model} specifies the system model of the cooperative virtual-MIMO system. Codebook design methods and the corresponding complexity analyses are investigated in Section \ref{section Design}. The effects of CFO and how they are overcome are illustrated in Section \ref{section_CFO}. Section \ref{section results} shows the simulation results and Section \ref{section conclusion} concludes the paper.

\section{Virtual-MIMO System with CF Cooperation}
\label{System Model}
\subsection{Channel Model}

\begin{figure}[!ht]
\centering
\includegraphics[width=15cm]{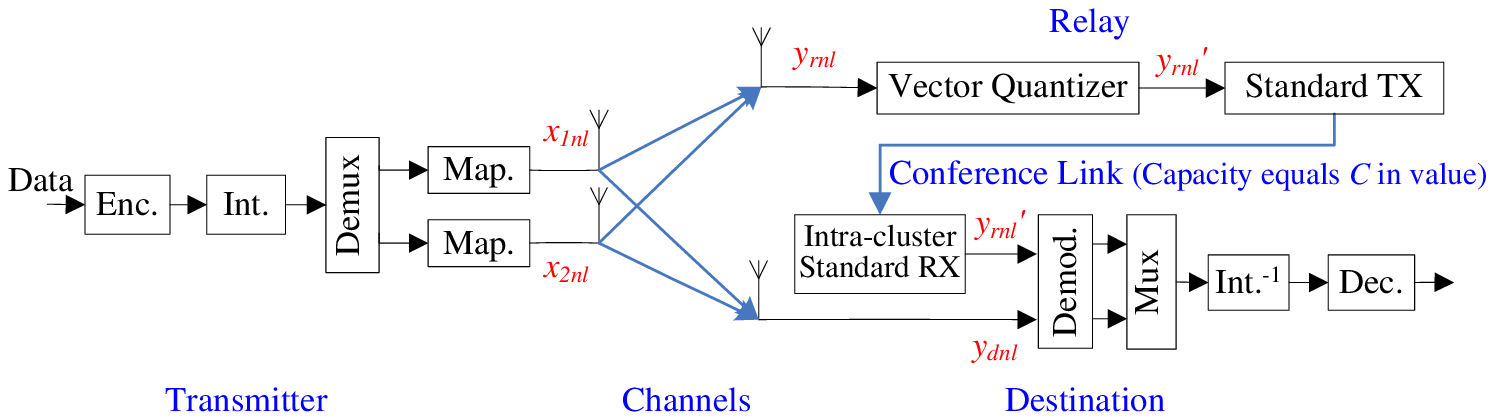}
\caption{System model of the cooperative virtual-MIMO system. (TX and RX stand for the transmitter and receiver.)
} \label{fig_system}
\end{figure}

Consider a cooperative virtual-MIMO network with one remote $N_{t}$-antenna transmitter sending information to $N_{r}$ collocated single-antenna receivers, as shown in Fig. \ref{fig_system}.
Since our research focuses on investigating a practical implementation of the CF cooperation for virtual-MIMO systems,  we start with a simple configuration with $N_{t}\!=\!N_{r}\!=\!2$. At the transmitter, the information bits are encoded through a rate-$R_{b}$ linear binary convolutional encoder.
We implement the BICM technique to provide FEC.
Thus the coded bits are interleaved through a random bit interleaver (int.).
In each steam after the demultiplexer (or demux), groups of $m$ bits are mapped onto complex data symbols via Gray-coded $2^{m}$ $=\!\!M$-ary PSK or QAM modulation.
Note that the system studied here is not limited to BICM, other FEC coding schemes, such as Turbo coding or LDPC coding, could also be employed according to different application requirements.

Since the transmitter is some distance from the receiver group,
a block fading channel model with $N$ Rayleigh fading blocks is assumed here: each block having length $L$ symbol periods.
When we consider a single time instance $l$ for the $n$th channel, the channel model is given by:
\begin{eqnarray}
\left[ {\begin{array}{*{20}c}\vspace{-5pt}
   {y_{rnl}}  \\\vspace{-5pt}
   {y_{dnl}}  \\
\end{array}} \right]
\!=\!{\emph{\textbf{H}}}_n {\textbf{\emph{x}}}_{nl}\!+\! {\textbf{\emph{w}}}_{nl},
\text{ with   }  {\emph{\textbf{H}}}_n \!=\! \left[\!\!  {\begin{array}{*{20}c}\vspace{-7pt}
   {h_{1n} } \!&\! {h_{2n} }  \\\vspace{-5pt}
   {h_{3n} } \!&\! {h_{4n} }  \\
\end{array}} \!\!\right]\!,
\label{eq_channel model}
\end{eqnarray}
where ${\emph{\textbf{H}}}_n$ denotes the $n$th block fading channel matrix, where each $h_{in} (i\!\in\![1,...,4])$ is independent and identically distributed (i.i.d.).
Without loss of generality, we assume normalized Rayleigh fading, i.e. $\mathbb{E}[|h_{in}|^2]\!=\!1$.
The complex scalars $y_{rnl}$ and $y_{dnl}$ are the signals at the relay and destination receivers.
We also define the vector ${\textbf{\emph{x}}}_{nl}\!=\![x_{1nl}, x_{2nl}]^{\text{T}}$ and ${\textbf{\emph{w}}}_{nl}\!=\![w_{1nl}, w_{2nl}]^{\text{T}}$, where the noise samples $w_{i'nl}\!\sim\! \mathcal{CN}(0,N_{0})$.
The average transmitted power per symbol is  $\mathbb{E}[|x_{i'nl}|^2]\!=\!E_s/N_t$. We normalize the total power $E_s$ to unity, and the corresponding power per bit is $E_b\!=\!E_s/(mR_{b})$. We assume that perfect CSI is available at the receivers only.

As the destination and the relay receivers are closely spaced, it is reasonable to expect that a high capacity communication link with high reliability can be formed between them.
Hence, as also considered in \cite{Dabora08} and \cite{Ng2006}, we assume the two receivers cooperate by way of an error-free conference link, as shown in Fig. \ref{fig_system}. We consider one-shot conference cooperation \cite{Ng2006}\cite{Draper03}, which requires the destination to decode the signal sent over the conference link. In practice,
the short-range conference link is realized via an orthogonal channel (i.e. a different frequency band) to the transmitter array.
Compared with the long data channel ${\emph{\textbf{H}}}_n$,
the orthogonal conference link is short-range allowing much higher rate transmission, permitting it to be reused many times over the coverage area of the long range link.

\subsection{Compress-and-Forward Cooperation}
\label{section CF}

The conference link enables cooperation, and CF could reasonably be used since it provides a higher rate when the relay is closer to the destination\cite{Host-Madsen1}. To perform CF cooperation, a standard source coding technique is employed for practical considerations. The reason why we do not employ the Wyner-Ziv (WZ) coding technique is that the virtual-MIMO system with multiple antennas at the transmitter has the feature that $y_{rnl}$ and $y_{dnl}$ are not highly correlated. The WZ coding technique therefore does not improve the performance significantly\cite{Wittneben}\cite{Jiang2}, but instead introduces extra complexity. Since standard source coding is simpler and also performs well in practical scenarios, we choose to implement it at the relay.
That is, the relay is equipped with a vector quantizer (VQ), as shown in Fig. \ref{fig_system}. We define the quantization rate (i.e. source coding rate) as $C$, which is measured in bits per compressed sample. As the two receivers cooperate by way of an error-free conference link, the link capacity is equal to the source coding rate $C$.
A compressed version of the signal ${y}'_{rnl}$ could be modeled by adding to $y_{rnl}$ an i.i.d. complex Gaussian noise\cite{JiangGC10},
\begin{equation}
y'_{rnl}=  {y_{rnl}+w_{cnl}},
\label{eq_ynlN}
\end{equation}
where $w_{cnl}$ is the compression noise with variance $\sigma_{cn}^{2}$, i.e. $w_{cnl}\sim \mathcal{CN}(0,\sigma_{cn}^{2})$.
A lower bound on the compression noise variance given in \cite{Kim}, which is computed by using Shannon's rate-distortion theory, could be extended to our virtual-MIMO system, i.e.,
\begin{equation}
\bar{\sigma}_{cn}^{2}=\dfrac{\mathbb{E}[|y_{rnl}|^{2}]}{2^{C}-1}=\dfrac{N_0+\frac{1}{2}|h_{1n}|^{2}+\frac{1}{2}|h_{2n}|^{2}}{2^{C}-1}.
\label{eq_cnbound}
\end{equation}

The general structure of the destination receiver is shown in Fig. \ref{fig_system}.
It receives signals $y_{dnl}$ from the transmitter, and observes signals from the relay via an intra-cluster receiver, written as ${y}'_{rnl}$ because of the error-free conference link. We denote the received signals at the destination as ${\emph{\textbf{y}}}_{nl}\!=\!\left[ {{y}'_{rnl} \;y_{dnl} } \right]^{\text{T}} $.
The destination requires knowledge of $h_{1n}$, $h_{2n}$, and ${\sigma}_{cn}^{2}$ sent from the relay.
Assuming $w_{cnl}$ is i.i.d. complex Gaussian,  $y'_{rnl}$ could then be scaled so that $y'_{rnl}$ and $y_{dnl}$ experience the same power level of additive Gaussian noise,
\begin{eqnarray}
\tilde{\emph{\textbf{y}}}_{nl}=\left[ {\tilde{y}'_{rnl} \;\;y_{dnl} } \right]^{\text{T}}  = \widetilde{\emph{\textbf{H}}}_{n}\textbf{\textit{x}}_{nl}  + \left[ {\tilde w_{1nl} \;\;w_{2nl} } \right]^{\text{T}},\label{eq_ycnl}\\
\text{with  }\;\;
\widetilde{\emph{\textbf{H}}}_{n}   \buildrel \Delta \over
= \! \left[ {\begin{array}{*{20}c}\vspace{-6pt}
   {\sqrt {\eta _{n} } h_{1n} } & {\sqrt {\eta _{n} } h_{2n} }  \\\vspace{-6pt}
   {h_{3n} } & {h_{4n} }  \\
\end{array}} \right],
\; \eta _{n}  \!   \buildrel \Delta \over =  \! \frac{{N_0 }}{{N_0  + \sigma _{cn} ^2 }}.
\label{eq_Hcn}
\end{eqnarray}
Here $\tilde w_{1nl}\sim$ i.i.d. $\mathcal{CN}(0,N_{0})$, $\eta _{n} $ is the degradation factor due to the compression noise, and  $\tilde{y}'_{rnl}  = {\sqrt {\eta _{n} } y'_{rnl} }$.
The scaled channel matrix $\widetilde{\emph{\textbf{H}}}_{n}$ will help the destination to mitigate the effects of the compression noise.
Next the destination performs joint maximum-likelihood (ML) demodulation of $\tilde{\emph{\textbf{y}}}_{nl}$, computing the log-likelihood ratio (LLR) for each coded bit.
Finally the decoder accepts the deinterleaved LLRs of all coded bits and employs a soft-input Viterbi algorithm to decode the signals.
Thus with help from the relay, the single-antenna destination receives signals from two transmit antennas.
As shown in (\ref{eq_cnbound})-(\ref{eq_Hcn}), with fixed SNR, a high source coding rate $C$ will result in $\sigma_{cn}^{2}$ decreasing to 0, and then $\widetilde{\emph{\textbf{H}}}_{n}$ tends towards $\emph{\textbf{H}}_{n}$ in value.
A good quality compression scheme with a high value of $C$ will allow the destination to use $y_{rnl}'$ for MIMO decoding, and enable the virtual-MIMO to achieve almost ideal MIMO performance.

\section{Vector Quantization Design at the Relay}
\label{section Design}

To perform CF cooperation, a standard source coding technique (i.e. VQ) is employed, shown in Fig. \ref{fig_system}.
The key tasks of the relay thus include constructing a good codebook, quantizing the received signals, and forwarding the compressed signals ${y}'_{rnl}$ to the destination. The codebook design techniques and the corresponding complexities will be analysed in this section.

\subsection{Codebook Design}
\label{subsection codebook}

The codebook design at the relay is based on the number of codebook vectors which equals $2^{C}$ and requires knowledge of the noise-free constellation. Note that, besides signal symbols, some control information such as the modulation type is also transmitted on control channels in practice. It is reasonable to expect the relay could construct the noise-free constellation of the received signals, i.e. the constellation of $h_{1n}x_{1nl}\!+\!h_{2n}x_{2nl}$, denoted by $y_{rnl}^{c}$. The codebook design which uses $y_{rnl}^{c}$ as the training vectors, will be simple and efficient.
In this paper, Voronoi VQ, and TSVQ\cite{Gersho} are employed at the relay. Voronoi VQ is considered first, as it has the advantage that the codebook is optimal in the sense of minimising average distortion. To design it, the Linde-Buzo-Gray (LBG) algorithm which is based on the iterative use of codebook modification, is used\cite{Jiang}. However, Voronoi VQ implies high computational and search complexity, especially for high-order modulations and large $C$, as will be shown in Section \ref{subsection Complexity}. 

\begin{figure}[!ht]
\centering
\includegraphics[width=10.5cm]{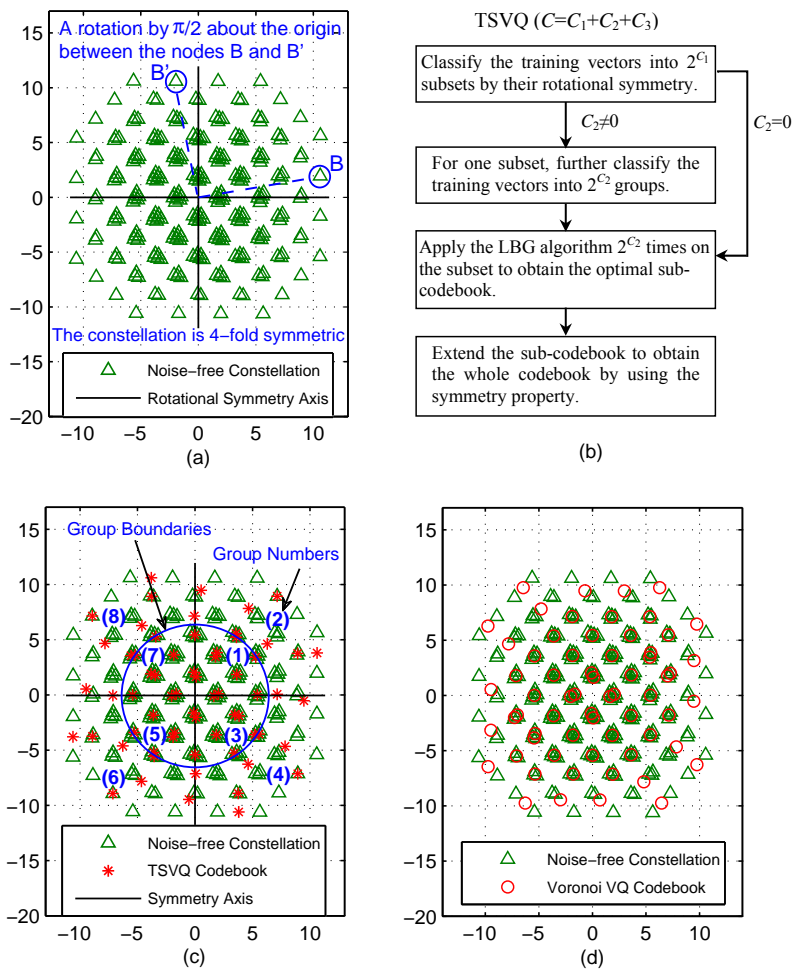}
 \vspace{-10pt}
\caption{Noise-free constellation of the received signals and the codebook designed at the relay with $C$= 6 bits/sample. (The rotational symmetry of the noise-free constellation is shown in (a); The conceptual diagram of the design process for TSVQ is in (b); The TSVQ codebook with $C_{1}$= 2 bits/sample and $C_{2}$= 1 bit/sample is shown in (c); The Voronoi VQ codebook is in (d).)}
\label{fig_cons2}
\end{figure}

To reduce the complexity, we also consider TSVQ for high-order modulations. Unlike the Voronoi VQ which requires a global and exhaustive search, the selection of the codebook vector in TSVQ can be divided into several stages and allows different design methods for each.
Simulation results suggest that, a combination of two path high-order rotationally symmetric constellations is also rotationally symmetric. For example, as shown in Fig. \ref{fig_cons2} (a), the combination of two 16QAM constellations from two paths, $y_{rnl}^{c}$, is four-fold rotationally symmetric: the origin is the centre of rotation, and $\pi/2$ is the angle of rotation. That is, the vectors of $y_{rnl}^{c}$ look the same after $\pi/2$ rotations are applied. 
The first stage quantization of TSVQ could therefore come from the rotational classification, i.e. classifying $y_{rnl}^{c}$ into some subsets based on the rotational symmetry, as illustrated in Fig. \ref{fig_cons2} (b). 
The sub-codebook designed for one subset would then be easily extended to the whole codebook, with the phase angles of the sub-codevectors changing by $\pi/2$ every time.
The final stage of the TSVQ could be determined by applying the LBG algorithm on the subsets to obtain optimal final-stage codebooks.
If there exist more than two stages, the middle stage could be resolved by using the LBG algorithm, or by classification according to some scheme  (e.g., magnitude and phase). The disadvantage of LBG for the middle stage is that it not only needs to store the entire training subsets corresponding to the sub-codebooks for further stage TSVQ design, but also implies a higher design complexity.
For practical considerations, the classification method is a better choice to design the middle stage of TSVQ. Here we denote the quantization rates for the three stages of TSVQ as $C_{1}$, $C_{2}$ and $C_{3}$ and we have $C=C_{1}+C_{2}+C_{3}$. 

Fig. \ref{fig_cons2} demonstrates a codebook design example where we suppose the two signals from the transmitters are 16QAM modulated. A set of $y_{rnl}^{c}$ for one 
choice of channel coefficients is shown in Fig. \ref{fig_cons2} (a). The points could be firstly divided into four subsets (i.e. $C_{1}=2 $ bits/sample) according to their rotational symmetry.
For one subset, we sort the constellation points in ascending order by their magnitudes and phases, and divide the subset into two groups in ascending order, i.e. $C_{2}=1 $ bit/sample.
After the two stages of TSVQ, we have 8 groups, labeled (1) - (8), as shown in Fig. \ref{fig_cons2} (c). Then the LBG algorithm is employed twice for group No. (1) and (2) to obtain the sub-codebook for one quadrant. The final stage quantization for TSVQ is thus completed by phase rotations. 
Compared to the Voronoi VQ used in Fig. \ref{fig_cons2} (d), the TSVQ we implemented here is much simpler to design and can achieve similar performance, as will be shown in Section \ref{section results}.
The illustrative example we present here uses $C_{2}=1$ bit/sample; 
A higher value of $C_{2}$ could be considered to further decrease the computational complexity, or one can set $C_{2}=0$ bits/sample for a high-accuracy quantization. 

Note that depending on the source coding rate $C$, there may not be enough codebook vectors to cover all noise-free constellation points. This is evident in Fig.~\ref{fig_cons2}~(d) for example, where one codebook vector lies among several adjacent modulation symbols. We rely then on the MIMO decoding at the destination to resolve ambiguities and correctly decode the data.

\subsection{Complexity Analysis}
\label{subsection Complexity}

As described above, both the Voronoi VQ and TSVQ use the LBG algorithm, but with a different size of training sequence. The codebook design complexity therefore comes from the computational complexity of LBG. The computational time for the LBG algorithm is\cite{Shanbehzadeh}:
\vspace{-2pt}
\begin{equation}
\vspace{-2pt}
T_{\text{LBG}}=I_{s}2^{C}QT_{d}+I_{s}(2^{C}-1)QT_{c},
\end{equation}
where $2^{C}$ is the codebook size, $I_{s}$ is the number of iterations,
and $Q$ is the number of training vectors. $T_{d}$ and $T_{c}$ denote the computational time for one distortion value and comparing two distortion values, respectively.
For the Voronoi VQ which implements the LBG algorithm on the whole noise-free constellation $y_{rnl}^{c}$, we have $Q=M^{2}=2^{2m}$. Since $I_{s}\leq Q/2^{C}$\cite{Shanbehzadeh}, we obtain:
\vspace{-3pt}
\begin{equation}
\vspace{-2pt}
T_{\text{d,Voronoi}}= T_{\text{LBG}}\leq 2^{4m}T_{d}+2^{4m}\frac{(2^{C}-1)}{2^{C}}T_{c}.
\label{eq_TVoronoi}
\end{equation}
For the TSVQ we implemented, the computational time is the summation of three stages,
\begin{eqnarray}
T_{\text{d,TSVQ}}\leq
{\left[\frac{2^{2m}}{4}T_{d} +\frac{2^{2m}}{8}\!\left(\!\frac{2^{2m}}{4}\!-\!1\!\right)\!T_{c}\right]  \text{sgn}(C_2) +2^{C_2}\!\left[\frac{2^{4m}}{2^{2(C_{1}+C_{2})}}T_{d}+\frac{2^{4m}}{2^{2(C_{1}+C_{2})}}\frac{2^{C_{3}}-1}{2^{C_{3}}}T_{c}\right] },
\label{eq_TTSVQ}
\end{eqnarray}
where $ \text{sgn}(C_2)$ denotes  the signum function of $C_2$.  
A justification for (\ref{eq_TTSVQ}) is as follows: In accordance with the multistage TSVQ we designed in Section \ref{subsection codebook}, we get $C_{1}\!=\!2$ bits/sample for the first stage of TSVQ.
Then we firstly consider the case $C_{2}\!>\!0$.
In one quadrant, there are $2^{2m}/{4}$ vectors, the distortion values of which are to be computed and sorted to obtain several separate groups for the second stage quantization. Here the selection sort algorithm is considered, requiring $\frac{2^{2m}}{4}(\frac{2^{2m}}{4}- 1) / 2$ comparison operations. For the final stage, the LBG algorithm is implemented $2^{C_2}$ times for the $2^{C_2}$ groups in one quadrant. In one group, there are $Q=2^{2m}/2^{(C_{1}+C_{2})}$ training vectors and the codebook size is $2^{C_{3}}$. So we obtain an upper bound of $T_{\text{d,TSVQ}}$ as shown in (\ref{eq_TTSVQ}). For the case $C_{2}\!=\!0$, it is obvious that the computational complexity comes from the LBG algorithm used in the final stage quantization where $2^{2m}/{2^{C_{1}}}$ training vectors are considered.

For example, when we consider $C_{1}=2$ bits/sample and $C_{2}=1$ bit/sample, we have $T_{\text{d,Voronoi}}=\mathcal{O}(2^{4m}(T_{d}+T_{c}))$ and $T_{\text{d,TSVQ}}=\mathcal{O}(2^{4m-5}T_{d}+2^{4m-4}T_{c})$, for a high-order modulation (e.g 16QAM) and a large value of $C$. That is, compared with the Voronoi VQ, TSVQ decreases the computational complexity for the codebook design significantly.

As to the symbol encoding, TSVQ will also allow a faster codebook search. Specifically, the encoding algorithm for a Voronoi VQ can be viewed as an exhaustive search algorithm. For a codebook of size $2^{C}$, the codevector selection for one symbol requires $2^{C}$ distortion evaluations and $2^{C}-1$ comparisons. The required time to search the codebook for one symbol is shown as:
\begin{equation}
T_{\text{s,Voronoi}}=2^{C}T_{d}+(2^{C}-1)T_{c}.
\label{eq_TsVo}
\end{equation}
For the TSVQ we designed, the search procedure includes two steps: finding out an appropriate group, and performing a full search on the group. Thus the search time of TSVQ is,
\begin{equation}
T_{\text{s,TSVQ}}=(C_{2}+2^{C_{3}})T_{d}+(C_{2}+2^{C_{3}}-1)T_{c}.
\label{eq_TsTSVQ}
\end{equation}

Therefore, compared with the Voronoi VQ, the multistage TSVQ not only decreases the computational complexity for the codebook design, but also allows a faster codebook search for the encoding. Since the multistage TSVQ can also achieve a good performance, it is a better choice to enable CF cooperation in practice.

\section{Effects of Carrier Frequency Offset}
\label{section_CFO}

With CF cooperation, the destination may expect traditional MIMO benefits in the virtual-MIMO system. However, due to the different oscillator frequencies at the source, relay and destination receivers, carrier frequency offsets (CFOs) occur\cite{Yilmaz}\cite{Zhang}. It will cause severe performance degradation, since CFOs between the transmitter-to-relay and the transmitter-to-destination channels can result in inter-block interference and distort the correlation properties of the signals at the final destination. This section thus focuses on the effects of CFO in the virtual-MIMO system. A clock synchronization and joint CFO estimation scheme is then proposed.

\subsection{CFO Estimation}
\label{subsection_CFO Estimation}

CFOs cause continuous phase rotations of the corresponding signals, and thus impair the detection performance. To alleviate this effect, the CFOs have to be estimated and then compensated at the receivers. The estimation of CFO in a MIMO system has been investigated in the literature \cite{McKeown,Deng,Yao}. The methods in \cite{McKeown} and \cite{Deng} are pilot aided requiring training sequences,
whereas \cite{Yao} develops a blind CFO estimation technique.
The CFO estimator in \cite{McKeown} is based on the measurement of the phase shift between consecutive channel estimation sequences. As it is simple and effective in practice, we firstly implement it at the relay and the destination as a baseline case.

Since $N_t$ antennas at the transmitter are fed with the same clock, frequency offsets at the relay and destination are defined as $f_{rn}$ and $f_{dn}$, respectively.
We let $[\emph{\textbf{u}}_{nl}, \emph{\textbf{u}}_{n(l+1)}]$ denote the channel estimation sequence, which occupies two symbol periods as $N_t\!=\!2$, and $\emph{\textbf{u}}_{nl}$ is an $N_t\!\times\!1$ vector transmitted through the channel matrix $\emph{\textbf{H}}_{n}$\cite{McKeown}. Channel estimation sequences are transmitted continuously on a dedicated pilot channel. The received signals at the relay and destination receivers corresponding to the channel estimation sequence $\emph{\textbf{u}}_{nl}$ at time $t_{nl}$ is then given by \cite{McKeown}\cite{Weng2007},
\begin{eqnarray}
\left[\! {\begin{array}{*{20}c}
   {y_{rnl}^u}  \\
   {y_{dnl}^u}  \\
\end{array}} \! \! \right] \!=\!
\left[\!\! {\begin{array}{*{20}c}
   {e^{j(2\pi{f}_{rn}t_{nl})} } \!&\! {0 }  \\
   {0 } \!&\! {e^{j(2\pi{f}_{dn}t_{nl})} }  \\
\end{array}} \!\!\right]
{\emph{\textbf{H}}}_n {\textbf{\emph{u}}}_{nl}+ {\textbf{\emph{w}}}_{nl}^u,
\label{eq_channelU}
\end{eqnarray}
where ${\textbf{\emph{w}}}_{nl}^u$ is the $N_t\!\times\!1$ complex Gaussian noise vector, with the noise samples $\sim\! \mathcal{CN}(0,N_{0})$. The equation (11) represents the phase rotations of the received signals caused by CFOs. According to \cite{McKeown}, we let the matrix ${\textbf{\emph{P}}}_{rn1}$ be formed by collecting the received signals at the relay corresponding to $[\emph{\textbf{u}}_{nl}, \emph{\textbf{u}}_{n(l+1)}]$, i.e. ${\textbf{\emph{P}}}_{rn1}=[{y_{rnl}^u}, {y_{rn(l+1)}^u}]$, and let ${\textbf{\emph{P}}}_{rn2}$ be defined as containing that pertaining to the next transmission of the channel estimation sequence. Likewise, ${\textbf{\emph{P}}}_{dn1}$ and ${\textbf{\emph{P}}}_{dn2}$ are defined for the destination. Then CFOs $f_{rn}$ and $f_{dn}$ can be estimated as,
\begin{eqnarray}
\tilde{f}_{rn}=\arg[{\textbf{\emph{P}}}_{\!rn2}\cdot{\textbf{\emph{P}}}_{\!rn1}^{\;\dagger}]/(2\pi\cdotp2t_s), \label{eq_CFOestJr}\\
\tilde{f}_{dn}=\arg[{\textbf{\emph{P}}}_{\!dn2}\cdot{\textbf{\emph{P}}}_{\!dn1}^{\;\dagger}]/(2\pi\cdotp2t_s),
\label{eq_CFOestJd}
\end{eqnarray}
where $\arg[\cdotp]$ denotes the angle operation, $t_s$ denotes the symbol interval, and $\{\cdot\}^{\dagger}$ represents the Hermitian transpose.

\subsection{Clock Synchronization and Joint CFO Estimation Scheme}
\label{subsection_CFO Synch}

The basic motivation for this proposed scheme is the fact that CFO causes severe performance degradation for the $2\times2$ virtual MIMO system.
It is shown in \cite{McKeown} and \cite{Zhang}
that a higher number of transmit and receive antennas, e.g. $4\times4$ and $8\times8$ MIMO systems, leads to significant improvement in the accuracy of the CFO estimate. However, for the simpler $2\times2$ MIMO or $2\times1$ MISO systems, residual CFO will still impair the detection performance. Meanwhile, cooperative virtual-MIMO systems also suffer performance degradation as compared to conventional MIMO where all the antennas at the receiver are fed with the same clock. In virtual-MIMO systems, cooperating antennas which are running with different clocks require to estimate and compensate their CFOs independently, as shown in equations (\ref{eq_CFOestJr}) and (\ref{eq_CFOestJd}). Their residual CFOs, i.e. ($\tilde{f}_{rn}\!-\!{f}_{rn}$) and ($\tilde{f}_{dn}\!-\!{f}_{dn}$), are different and independent. Since the operation at the relay does not cause extra phase rotations, the different residual CFOs will then distort the correlation properties of the received signals at the destination, so that the system performance is impaired. To mitigate this effect, the use of frequency-domain equalization has been proposed in cooperative systems\cite{Yilmaz}\cite{Yadav}, but only for a single-antenna transmitter.
For our virtual-MIMO system which has multiple antennas transmitting different data,
we propose to maintain clock synchronization across the receivers, and then jointly estimate CFO between both terminals.

The proposed scheme involves two steps. Clock synchronization is the first step for providing a common notion of time across the relay and the destination.
State of the art synchronization algorithms are described in \cite{Wu}, assuming that the transmissions for delivering time information are line of sight.
To estimate the frequency difference between the receivers, a two-way message exchange (which is a classical timing message signaling approach) could be implemented.
Consider the destination as the reference node, so that the relay needs to synchronize with the destination. It requires timing messages to be exchanged several times to achieve a certain synchronization accuracy.
The stable nature of the error-free conference link between the relay and the destination is helpful for this synchronization process.
In practice, the conference link is short range, and
it is reasonable to expect that in many cases the conference link is a reliable line-of-sight link with high bandwidth, which may support frequent timing message exchanges.
Hence, on the stable short-range conference link, the relay and the destination could maintain clock synchronization.

\begin{figure}[!ht]
 \vspace{50pt}
\centering
\includegraphics[width=12cm]{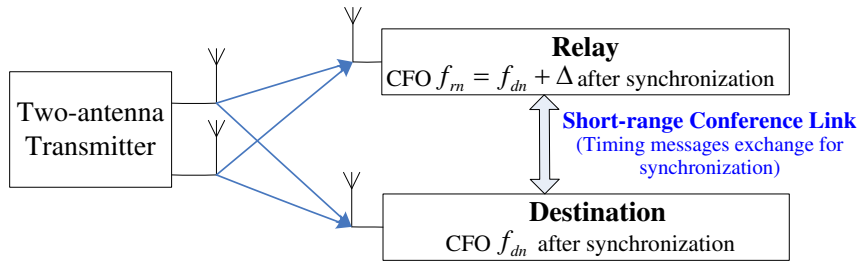}
\caption{Carrier frequency offsets at the relay and the destination.}
\label{fig_CFO}
\end{figure}

The clock synchronization approach studied here focuses specifically on frequency locking, as the effect of frequency offset is the main reason why clock offsets drift over time\cite{Noh}. Compensating the frequency offset guarantees long-term reliability of synchronization, so that re-synchronization only need to be preformed for each $\emph{\textbf{H}}_{n}$.
Without loss of generality, after the clock synchronization, frequency offsets at the relay and the destination are modelled by (as shown in Fig.\ref{fig_CFO}),
\begin{eqnarray}
{f}_{rn}={f}_{dn}+\Delta_{n},
\label{eq_CFO}
\end{eqnarray}
where $\Delta_{n}$ denotes the frequency synchronization error, which is determined by the synchronization algorithm, the delays in timing message delivery, and the number of observations of timing messages.
Efficient algorithms relying on two-way message exchanges have been reported in \cite{Wu}. For algorithms that do not compensate the frequency offsets, such as the timing-sync protocol for sensor networks (TPSN), synchronization has to be performed more frequently to maintain the required accuracy. The algorithm in \cite{Noh} computes and corrects the frequency offsets, and thus serves as a good candidate for our system. It is reported that the frequency synchronization error decreases as the number of timing messages exchanged increases.

The second step of the proposed scheme is to perform joint CFO estimation between the relay and the destination,
to obtain the benefit of MIMO CFO estimation.
Specifically, the relay computes $J_{rn}={\textbf{\emph{P}}}_{rn2}\cdot{\textbf{\emph{P}}}_{rn1}^{\;\dagger}$ and transmits it to the destination via the conference link. The destination receives $J_{dn}={\textbf{\emph{P}}}_{dn2}\cdot{\textbf{\emph{P}}}_{dn1}^{\;\dagger}$ and estimates $f_{dn}$ based on $J_{rn}$ and $J_{dn}$,
\begin{eqnarray}
\tilde{f}_{dn}=\arg[J_{rn}+J_{dn}]/(2\pi\cdotp2t_s).
\label{eq_CFOest}
\end{eqnarray}
The estimated CFO is then shared with the relay, i.e. $\tilde{f}_{rn}=\tilde{f}_{dn}$, to help the relay compensate its CFO ${f}_{rn}$ before vector quantization. Here we compare $\tilde{f}_{dn}$ against the estimated CFO of the corresponding MIMO system which is denoted by $\tilde{f}_{n}$.
Given a specific channel condition, when the synchronization error $\Delta_{n}= 0$, we have $\tilde{f}_{dn}$ computed using the joint CFO estimation scheme equals $\tilde{f}_{n}$ in value, and therefore the benefits of MIMO CFO estimation are exploited.
A brief proof is as follows: For the MIMO system, the received signals corresponding to  $\emph{\textbf{u}}_{nl}$ is given by ${\emph{\textbf{y}}}^{u}_{nl} \!=\!
{\emph{\textbf{H}}}_n {\textbf{\emph{u}}}_{nl}e^{j(2\pi{f}_{n}t_{nl})}+ {\textbf{\emph{w}}}_{nl}^u$.
According to \cite{McKeown}, ${\textbf{\emph{P}}}_{n1}$ is defined as $[{{\emph{\textbf{y}}}_{nl}^u}, {{\emph{\textbf{y}}}_{n(l+1)}^u}]$, and ${\textbf{\emph{P}}}_{n2}$ is for next transmission of the channel estimation sequence. Then we get a square matrix ${\textbf{\emph{J}}}_{n}={\textbf{\emph{P}}}_{n2}\cdotp{\textbf{\emph{P}}}_{n1}^{\;\dagger}$, so that ${f}_{n}$ can be estimated by,
\begin{eqnarray}
\tilde{f}_{n}=\arg[\;\text{tr}({\textbf{\emph{J}}}_{n})\;]/(2\pi\cdotp2t_s),
\label{eq_CFOestJdMIMO}
\end{eqnarray}
where $\text{tr}(\cdotp)$ denotes the trace operation.
If $\Delta_n=0$, we have ${\emph{\textbf{y}}}^{u}_{nl}=\left[ {{y}^{u}_{rnl} \;\;{y}^{u}_{dnl} } \right]^{\text{T}} $, so that,
\begin{eqnarray}
\text{tr}({\textbf{\emph{P}}}_{n2}\cdotp{\textbf{\emph{P}}}_{n1}^{\;\dagger})=\text{tr}\left(\left[\! {\begin{array}{*{20}c}
   {\textbf{\emph{P}}}_{rn2}  \\
   {\textbf{\emph{P}}}_{dn2}  \\
\end{array}} \! \! \right]\cdotp[{\textbf{\emph{P}}}_{rn1}^{\;\dagger}\;\;{\textbf{\emph{P}}}_{dn1}^{\;\dagger}]
\right)= {\textbf{\emph{P}}}_{rn2}\cdot{\textbf{\emph{P}}}_{rn1}^{\;\dagger}+{\textbf{\emph{P}}}_{dn2}\cdot{\textbf{\emph{P}}}_{dn1}^{\;\dagger}
\label{eq_CFOcom}
\end{eqnarray}
Substituting (\ref{eq_CFOcom}) into (\ref{eq_CFOestJdMIMO}), and comparing it with (\ref{eq_CFOest}), we finally get $\tilde{f}_{dn}=\tilde{f}_{rn}=\tilde{f}_{n}$.
The joint CFO estimation scheme therefore exploits spatial diversity of the MIMO receiver (when $\Delta_n=0$), and offers significant improvement compared to (\ref{eq_CFOestJd}).
The performance of the joint CFO estimation scheme is thus lower bounded by the perfect frequency-synchronized case.

For the case $\Delta_n\neq0$, at high SNR, $\tilde{f}_{dn}$ will tend toward ${f}_{dn}+ \Delta_n/2$,
\begin{eqnarray}
\tilde{f}_{rn}=\tilde{f}_{dn}={f}_{dn}+ \Delta_n/2={f}_{rn}-\Delta_n/2.
\label{eq_CFOhighSNR}
\end{eqnarray}
That is because the estimated $\tilde{f}_{dn}$ from (\ref{eq_CFOest}) is based on two CFO observations. When $\Delta_n\neq0$, with SNR increasing, $\tilde{f}_{dn}$ will equal the average of ${f}_{dn}$ and ${f}_{rn}$.
Thus for both the relay and the destination, the magnitude of CFO mismatch is $\Delta_n/2$. In this case, the joint CFO estimation scheme may still take advantage of the cooperative estimation, but the benefit will be reduced as $\Delta_n$ increases. Different values of $\Delta_n$ represent different degrees of synchronization across the relay and the destination.
The distribution of $\Delta_n$ is complicated, and is related to the synchronization algorithms, various delays in timing message delivery and the number of timing messages exchanged, as mentioned before.
In this paper, we consider a fixed value of $\Delta_n$ for all channel conditions and will drop its subscript $n$ in simulations (in Section \ref{section results}), in order to demonstrate the effects of the value of $\Delta$ on performance.

Using the two steps, we get the clock synchronization and joint CFO estimation scheme to counteract the performance degradations caused by CFOs.
Note that the joint CFO estimation studied here is based on McKeown's algorithm in \cite{McKeown}, but our approach can be applied to other estimation algorithms with suitable changes to the shared information to complete the joint CFO estimation operation. For example, if OFDM is used, the pilot symbols to be transmitted will be grouped into blocks, each block will be transformed by IDFT, and cyclic prefixes or some zeros will be added to the tail of each transformed block \cite{Weng2007}. 
The pilot blocks could be designed (as discussed in \cite{Zeng2007} and \cite{Chen2008}) and the CFO estimation algorithm will be selected accordingly, with the goal of providing a small estimation mean 
squared error. At the receiver side, using our proposed scheme, clock synchronization is performed first such that the CFOs at the relay and the destination are ${f}_{dn}+\Delta$ and ${f}_{dn}$, respectively. Then the relay and destination perform joint CFO estimation using the selected estimation algorithm (corresponding to the dedicated pilot design), and thus obtain the benefits of MIMO CFO estimation.
Note that, using OFDM, CFO estimation and compensation are performed before the serial-to-parallel conversion \cite{Weng2007}\cite{Zeng2007}.

\section{Numerical Results}
\label{section results}

In this section, we present the error performance of our cooperative virtual-MIMO system ($N_t \!=\! N_r \! = \! 2$). At the transmitter, a binary convolutional code is assumed with the generator polynomials $[133, 171]_{\text{octal}}$ ($R_b\!=\!1/2$). Gray-labeled QPSK or 16QAM modulation are considered.
The channels between the transmitter and receivers are assumed to be i.i.d normalized block Rayleigh fading, with $10^6$ fading blocks and each block has 196 consecutive symbol periods.
The simulation results are obtained using the Monte Carlo method. We plot the bit error ratio (BER) or block error ratio (BLER) against the information bit SNR, i.e $E_b/N_0$.

\vspace{-0.5em}
\subsection{BER Evaluation of VQ Design}
\begin{figure}[!ht]
\centering
\includegraphics[width=11cm]{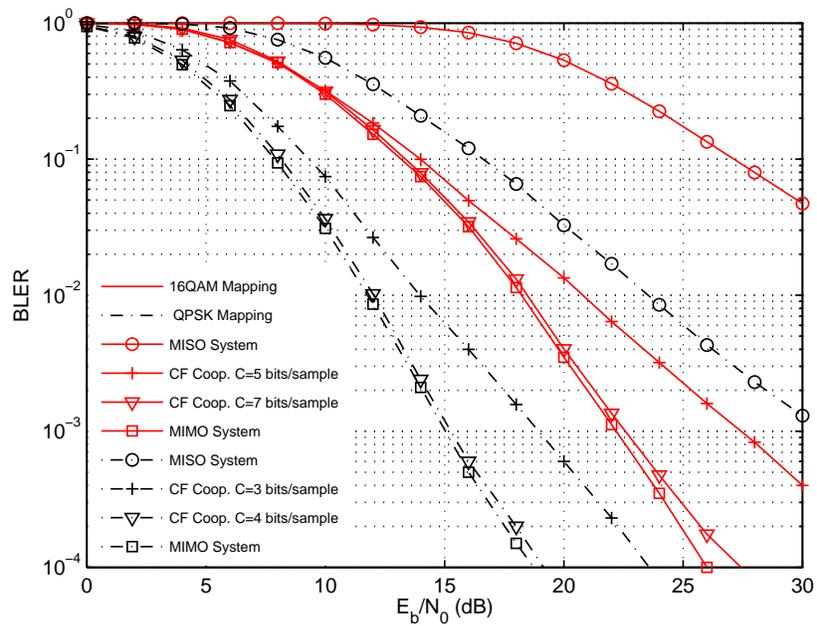}
\caption{BLER results of the cooperative virtual-MIMO system with Voronoi VQ at the relay and ML receiver at the destination. (Solid red curves correspond to 16QAM mapping, while dash-dotted black curves correspond to QPSK mapping.)}
\label{fig_16qam}
\end{figure}

\vspace{-0.3em}
The BLER performance of the cooperative virtual-MIMO system with the Voronoi VQ under various quantization rates, is shown in Fig. \ref{fig_16qam}. The BLERs are compared against the corresponding ideal MIMO system, and the non-cooperative MISO system.
Fig. \ref{fig_16qam} shows that,
with Voronoi VQ at the relay, $C\!=\!7$ bits/sample for 16QAM and $C=4$ bits/sample for QPSK modulation, will enable the system with CF cooperation to approach ideal MIMO performance.

\begin{figure}[!ht]
\centering
\includegraphics[width=11cm]{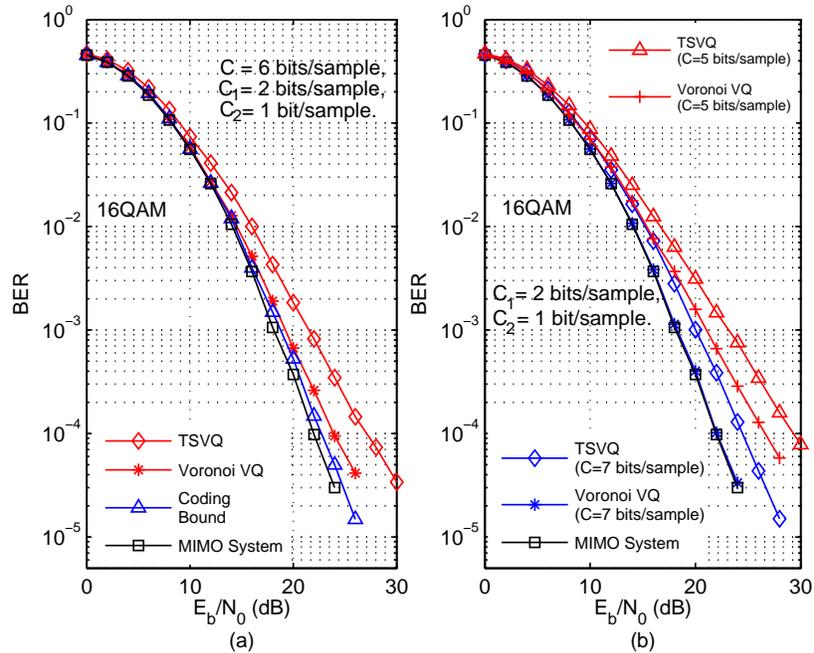}
\caption{BER performance of the cooperative virtual-MIMO system with TSVQ or Voronoi VQ at the relay. (Quantization rate 6 bits/sample is considered in (a), while 5 and 7 bits/sample are considered in (b).)}
\label{fig_16qamTSVQ}
\end{figure}

As mentioned above, to decrease the complexity, we propose to employ TSVQ to design the codebook at the relay.
When $C\!=\!$ 6 bits/sample, Fig. \ref{fig_16qamTSVQ} (a) shows the BER results of the TSVQ cooperative system which is set up in accordance with the example for 16QAM mapping in Section \ref{subsection codebook}. Its BER is compared against the performance of the system with the Shannon coding bound. According to equation (\ref{eq_ynlN}), the compression noise $w_{cnl}$ is considered for this kind of system, with the Shannon coding bound of the variance calculated via (\ref{eq_cnbound}).
As shown in this figure, the Voronoi VQ obtains performance which is quite close to the Shannon coding bound. Even though the TSVQ we designed here is suboptimal, it is much simpler to design and operate and can achieve error ratios comparable to the optimal but more complicated Voronoi VQ.
For a given quantization rate of 6 bits/sample, the Voronoi VQ requires a $\mathcal{O}(65536(T_{d}+T_{c}))$ computations which is much larger than that of TSVQ requiring $\mathcal{O}(2048T_{d}+4096T_{c})$ computations.
 
Additionally, considering the quantization rates 5 bits/sample and 7 bits/sample, we compare the performance of the Voronoi VQ and TSVQ in Fig. \ref{fig_16qamTSVQ} (b).
It can be seen that the BER of TSVQ for $C\!=\!6$ bits/sample in Fig. \ref{fig_16qamTSVQ} (a) performs almost the same as the Voronoi VQ for $C\!=\!5$ bits/sample in Fig. \ref{fig_16qamTSVQ} (b). According to (\ref{eq_TsVo}) and (\ref{eq_TsTSVQ}), that means the TSVQ with encoding complexity $(9T_{d}\!+\!8T_{c})$ is able to achieve the performance of the Voronoi VQ with $(32T_{d}\!+\!31T_{c})$  complexity. Also, the TSVQ for $C\!=\!7$ bits/sample performs similarly to the Voronoi VQ for $C\!=\!6$ bits/sample. That is, TSVQ could approach the performance of Voronoi VQ with roughly 4 times lower encoding complexity. TSVQ always requires a much lower computational complexity for the codebook design as well. Thus the TSVQ we implemented here is more efficient than the Voronoi VQ, and is a better choice for the CF cooperation to enable the virtual-MIMO system to achieve MIMO performance in practice.

\begin{figure}[!ht]
\centering
\includegraphics[width=11cm]{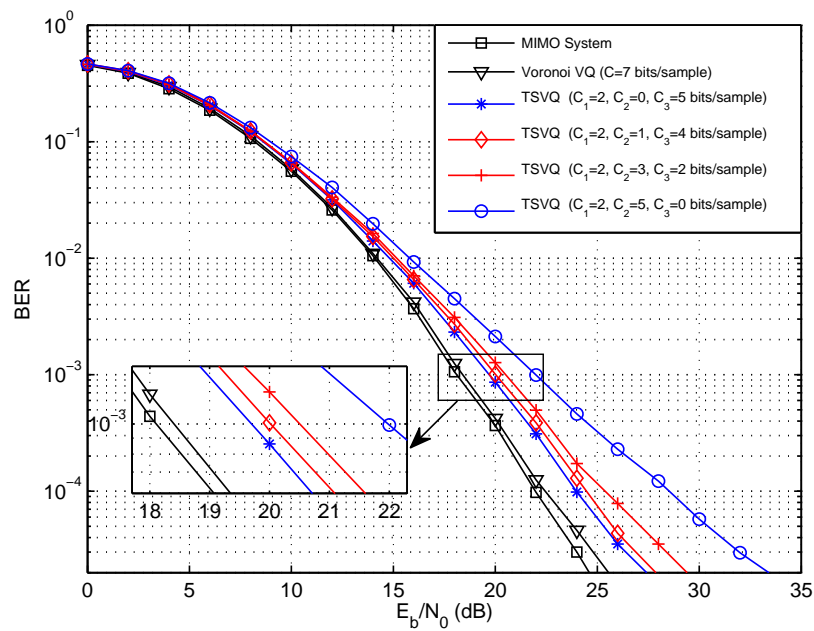}
\caption{BER comparison between the Voronoi VQ and TSVQ in the virtual-MIMO system. ($C=$ 7 bits/sample for the Voronoi VQ, and various $C_2$ for TSVQ are considered.)}
\label{fig_16qamTSVQC7}
\end{figure}

The above TSVQ results use $C_2\!=\!1$ bit/sample accorded to the illustrative example in Section \ref{subsection codebook}. But $C_2$ is not limited to that value: a higher $C_{2}$ could be considered to further decrease the computational complexity, or $C_{2}\!=\!0$ bits/sample can be used for a high-accuracy quantization, since $C_3\!=C-C_1-C_2$. The comparison of various values of $C_2$ for the TSVQ cooperative system is shown in Fig. \ref{fig_16qamTSVQC7}. We assume a total quantization rate $C\!=\!7$ bits/sample for 16QAM mapping in this figure. The corresponding design and encoding complexities at a BER of $10^{-3}$, are shown in TABLE \ref{table_complexity}. Compared to the Voronoi VQ, the TSVQ with $C_{2}\!=\!0$ bits/sample can reduce the codebook design complexity to one sixteenth, and decrease the encoding complexity to one quarter, for a ${E_b}/{N_0}$ performance penalty of only 1.3 dB.
Then as $C_2$ increases, both the design and encoding complexities reduce, but the BER performance becomes worse. Since both the second and third stage of TSVQ contribute to the complexities according to (\ref{eq_TTSVQ}) and (\ref{eq_TsTSVQ}), the complexity reduction is not proportional to the increase of $C_2$. Neither does the performance loss scale linearly with $C_2$ (i.e. the ${E_b}/{N_0}$ Loss). To achieve a favourable performance-complexity tradeoff, $C_2\!=\!0$ or 1 bit/sample is a good choice for this case.

\vspace{-0.5em}
\subsection{Effects of Carrier Frequency Offsets}
\vspace{-0.3em}
According to the 3GPP standards, the tolerance required for frequency accuracy is from $\pm$0.1 ppm (parts per million) to $\pm$0.25 ppm\cite{LTE}.
It translates to a CFO in the range up to $\pm$500 Hz at a carrier frequency of 2 GHz. For practical considerations, we assume the CFO at the destination follows a uniform distribution, i.e. ${f}_{dn}\sim$ U[-500, 500]~Hz, and again ${f}_{rn}={f}_{dn}+\Delta$. Moreover, a generic sub-frame structure defined in 3GPP LTE is adopted \cite{3GPP}: one sub-frame is made up of two 0.5 ms slots, each made of seven symbols, and thus the symbol interval $t_s=71.4 $ $\mu$s.
The frequency range that can be estimated is therefore $\pm$3500 Hz\cite{McKeown}, which can cover the maximum CFO assumed here.
Given a specific ${\emph{\textbf{H}}}_n$, the relay and the destination perform clock synchronization, and then jointly estimate and compensate their CFOs for each sub-frame.

\begin{figure}[!ht]
\centering
\includegraphics[width=13cm]{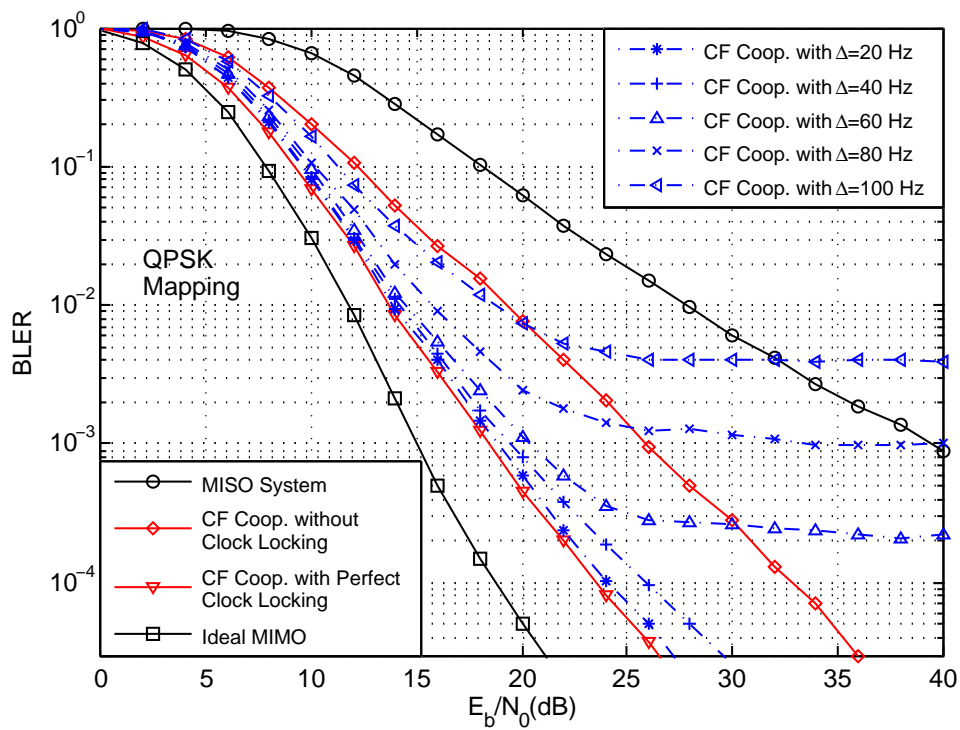}
\caption{BLER performance of the virtual-MIMO system with or without frequency synchronization for QPSK mapping. (Various degrees of synchronization and Voronoi VQ are considered.)}
\label{fig_CFOQPSK}
\end{figure}

The BLER performance of the $2\!\times\!2$ virtual-MIMO system with or without clock synchronization for QPSK modulation is shown in Fig. \ref{fig_CFOQPSK}. To focus on the effects of residual CFO, we firstly apply the Voronoi VQ with $C$= 4 bits/sample for cooperation. For the cooperative system without clock locking,
CFOs cause a drastic performance degradation compared to the ideal MIMO case.
In contrast, the proposed clock synchronization and joint CFO estimation scheme provides a significant performance advantage, which is lower bounded by the perfect clock locking case (i.e. $\Delta=$ 0 Hz).
Then a family of dash-dot curves shows the performance of various degrees of synchronization. For a large value of $\Delta$, there exists an error floor, which is caused by the residual CFO which equals $\Delta/2$ according to (\ref{eq_CFOhighSNR}).
When $\Delta \geq$ 100 Hz, joint CFO estimation is of little value, as it  does not provide a performance advantage compared to the case without clock locking. 
For a target BLER of $10^{-2}$, a synchronization error smaller than 60Hz provides a very good performance close to the lower bound.

\begin{figure}[!ht]
\centering
\includegraphics[width=13cm]{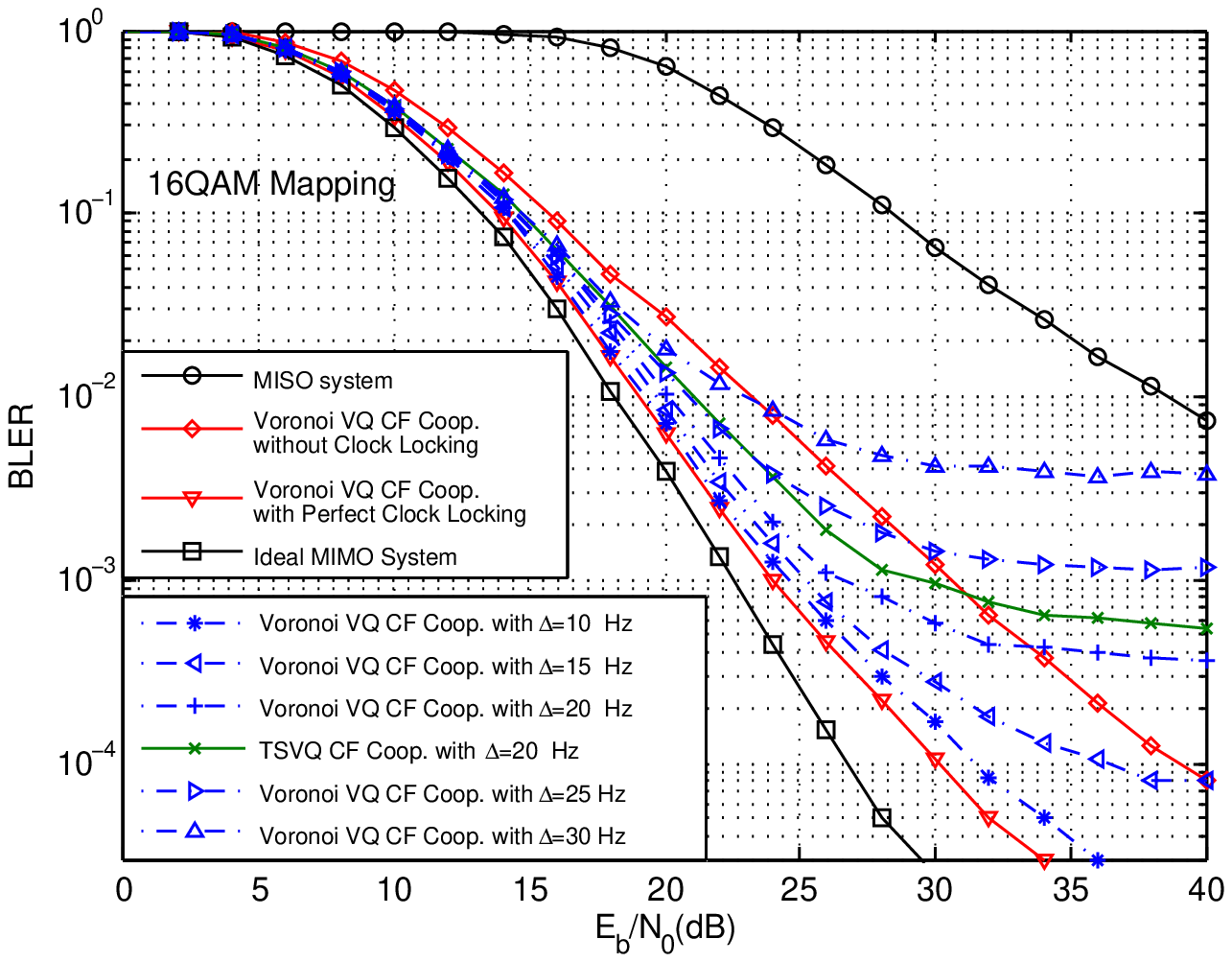}
\caption{BLER performance of the virtual-MIMO system with various degrees of frequency synchronization for 16QAM mapping. (Both Voronoi VQ and TSVQ are considered.)}
\label{fig_CFO16qam}
\end{figure}

A similar trend can be seen in Fig.\ref{fig_CFO16qam} for 16QAM mapping: Maintaining clock synchronization and jointly estimating CFO at the relay and the destination could provide a significant BER improvement which is close to the ideal MIMO case. Comparing the results for 16QAM and QPSK, it is obvious that 16QAM needs a higher level of synchronization because of its higher-order constellation, but QPSK could tolerate a larger synchronization error. For 16QAM, $\Delta\leq25$ Hz offers an acceptable performance, while $\Delta\leq20$ Hz could provide a very good performance for the virtual-MIMO system. Moreover, besides the Voronoi VQ with $C$= 7 bits/sample, this figure also applies the TSVQ we designed ($C_1$= 2, $C_2$= 1, $C_3$= 4 bits/sample) to enable CF cooperation for a specific case of the proposed scheme where $\Delta=20$ Hz is considered. The system performance with TSVQ follows a similar trend as the system with the Voronoi VQ.
For a target BLER of $10^{-2}$, TSVQ at $\Delta=20$ Hz performs almost the same as the Voronoi VQ at $\Delta=25$ Hz, but TSVQ has a much lower complexity. 

\section{Conclusions}
\label{section conclusion}
In this paper, a cooperative virtual-MIMO system using two transmit antennas that implements BICM transmission and CF cooperation among two receiving nodes was presented.
We proved that the CF cooperation using standard source coding at the relay could enable the virtual-MIMO system to achieve almost MIMO performance.
Then two codebook design algorithms were presented, Voronoi VQ and TSVQ, based on knowledge of the noise-free constellation. A comparison in terms of the codebook design complexity and encoding complexity was also presented.
We have shown that, compared to the Voronoi VQ, the TSVQ can reduce the codebook design complexity to less than one sixteenth, and decrease the encoding complexity to less than one quarter, for a performance penalty of only 1.3 - 1.6 dB.
A higher middle-stage quantization rate $C_{2}$ could be considered to further decrease the complexities, or $C_{2}\!=\!0$ bits/sample used for high-accuracy quantization.
In practice, the TSVQ is a better choice for CF cooperation to achieve a favourable performance-complexity tradeoff in the virtual-MIMO system.

Additionally, for practical considerations, we also investigated the effects of CFO, and demonstrated that CFO could lead to drastic performance degradation for the $2\times2$ virtual MIMO system. A scheme which maintains clock synchronization and jointly estimates CFO between the relay and the destination, is proposed to overcome the limitations of separate CFO estimation at the relay and destination. Simulation results showed that the proposed scheme provided a significant performance improvement. For a target BLER of $10^{-2}$, a synchronization error smaller than 60 Hz for QPSK and 20 Hz for 16QAM mapping, could offer good performance close to the case with perfect clock locking.

This paper dealt with two practical issues for the virtual-MIMO system with CF cooperation. We designed the efficient TSVQ as source coding technique at the relay, and proposed the clock synchronization and joint CFO estimation scheme so that the cooperation is not unduly sensitive to CFOs. The TSVQ is not limited to 16QAM mapping, the principles of which could easily be applied to multiple modulation types and quantization rates. Also, the clock synchronization and joint CFO estimation scheme could employ other estimation algorithms, besides McKeown's method. By extending to a wide range of applications, the virtual-MIMO system is therefore particularly valuable and attractive to some realistic wireless communication systems.
The extension to more cooperating terminals with more antennas is left as future work.

\section{Acknowledgements}
We would like to thank Prof. Norbert Goertz of the Vienna University of Technology for helpful comments and discussion.
We acknowledge the support of the Scottish Funding Council for the
Joint Research Institute with Edinburgh and Heriot-Watt
Universities, which is a part of the Edinburgh Research Partnership.

\ifCLASSOPTIONcaptionsoff
  \newpage
\fi

\begin{small}                     \end{small}

\newpage

\newpage

 \vspace{50pt}

\begin{table}[!ht]
\renewcommand{\arraystretch}{1.3}
\caption{Complexity and performance comparison between the Voronoi VQ and TSVQ.}
\label{table_complexity}
\centering
\begin{tabular}{|l|l|l|l|}
\hline
\raisebox{-0.5ex} {Design Algorithm} & \raisebox{-0.5ex} {Design} & \raisebox{-0.5ex} {Encoding} & \raisebox{-0.5ex}{${E_b}/{N_0}$ Loss}\\
\raisebox{0.5ex} {($C$ uses bit/sample unit)}& \raisebox{0.5ex} {Complexity}& \raisebox{0.5ex} {Complexity}&  \raisebox{0.5ex}{(BER=$10^{-3}$)}\\
\hline
\raisebox{-0.5ex}{Voronoi VQ ($C$= 7) }& \raisebox{-0.5ex}{$\mathcal{O}(65536(T_{d}+T_{c}))$} & \raisebox{-0.5ex}{$128T_{d}+127T_{c}$} & \raisebox{-0.5ex} {0 dB}\\[1ex]

{TSVQ ($C_1$= 2, $C_2$= 0, $C_3$= 5)} & $\mathcal{O}(4096(T_{d}+T_{c}))$ & $32T_{d}+31T_{c}$ & 1.3 dB\\[0.5ex]

{TSVQ ($C_1$= 2, $C_2$= 1, $C_3$= 4)} & $\mathcal{O}(2048T_{d}+4096T_{c})$ & $17T_{d}+16T_{c}$ & 1.6 dB\\[0.5ex]

{TSVQ ($C_1$= 2, $C_2$= 3, $C_3$= 2)} & $\mathcal{O}(512T_{d}+2048T_{c})$ & $5T_{d}+10T_{c}$ & 2.2 dB\\[0.5ex]

{TSVQ ($C_1$= 2, $C_2$= 5, $C_3$= 0)} & $\mathcal{O}(192T_{d}+2048T_{c})$ & $T_{d}+31T_{c}$ & 3.6 dB\\[0.5ex]
\hline
\end{tabular}
\end{table}

\begin{thebibliography}{1}

\bibitem{Gesbert}
Gesbert, D., Shafi, M., Shiu, D.-S., Smith, P., and Naguib, A.: `From theory to
  practice: an overview of {MIMO} space-time coded wireless systems',
  IEEE Journal on Selected Areas in Communications, 2003, 21, (3), pp. 281--302


\bibitem{Ng}
Ng, C.~T.~K., Jindal, N., Goldsmith, A., and Mitra, U.: `Capacity gain from
  two-transmitter and two-receiver cooperation', IEEE Trans. Inform.
  Theory, 2007, 53, (10), pp. 3822 -- 3827


\bibitem{Jiang_TVT1}
Jiang, J., Thompson, J. S.,  and Sun, H.: `A singular-value-based adaptive
modulation and cooperation scheme for virtual-MIMO systems', IEEE
Trans. Vehicular Technology, 2011, 60, (6), pp. 2495 -- 2504 


\bibitem{Liang}
Liang, Y., and Veeravalli, V.~V.: `Cooperative relay broadcast channels',
IEEE Trans. Inform. Theory, 2007, 53, (3), pp. 900 -- 928

\bibitem{Ding2010}
Ding, Z., Leung, K.K., Goeckel, D.L., and Towsley, D.: `Cooperative Transmission Protocols for Wireless Broadcast Channels', IEEE Trans. Wireless Communications, 2010, 9, (12), pp.3701 -- 3713


\bibitem{Caire}
Caire, G., Taricco, G., and Biglieri, E.: `Bit-interleaved coded modulation',
IEEE Trans. Inform. Theory, 1998, 44, (3), pp. 927 -- 946

\bibitem{Krikidis2008}
Krikidis, I., Thompson, J., McLaughlin, S., and Goertz, N.: `Optimization issues for cooperative amplify-and-forward systems over block-fading channels', IEEE Trans. on Vehicular Technology, 2008, 57, (5), pp. 2868 -- 2884


\bibitem{Host-Madsen1}
Host-Madsen, A., and Zhang, J.: `Capacity bounds and power allocation for
  wireless relay channels', IEEE Trans. Inform. Theory, 2005, 51, (6),
  pp. 2020 -- 2040

\bibitem{Qi2009}
Qi, Y., Hoshyar, R., and Tafazolli, R.: `A Novel Quantization Scheme in Compress-and-Forward Relay System',   Proc. IEEE VTC 2009 Spring, Barcelona, Spain, April 2009, pp. 1 -- 5.

\bibitem{Jiang}
Jiang, J., Thompson, J. S., Grant, P. M., and Goertz, N.: `Practical
  compress-and-forward cooperation for the classical relay network',
  Proc. European Signal Processing Conference, Glasgow, Scotland, {A}ug. 2009, pp. 2421 -- 2425
 

\bibitem{Jiang2}  
Jiang, J., Thompson, J. S., Sun, H., and Grant, P. M.: `Performance assessment of virtual multiple-input multiple-output systems with compress-and-forward cooperation',  IET Communications, 2012, 6, (11), pp. 1456 -- 1465 


\bibitem{McKeown}
McKeown, M., Cruickshank, D., Lindsay, I., Thompson, J. S., Farson, S., and Hu, Y.: `Carrier frequency offset estimation in {BLAST} {MIMO} systems', Electronics Letters, 2003, 39, (24), pp. 1752--1753 

\bibitem{Deng}
Deng, K., Tang, Y., Shao, S., and Li, S.: `Carrier frequency offset estimation for
  {MIMO} correlated fading channels', Proc. IEEE Wireless Communications
  and Networking Conference (WCNC), Kowloon, Hong Kong, {M}ar. 2007, pp.
  1035--1038

\bibitem{Yao}
Yao Y., and Giannakis, G.: `Blind carrier frequency offset estimation in {SISO},
  {MIMO}, and multiuser {OFDM} systems', IEEE Trans. Commun., 2005, 53,
  (1), pp. 173--183 
  
\bibitem{Weng2007}  
Weng L., Au, E. K. S., Chan, P. W. C., Murch, R. D., Cheng, R. S., Mow, W. H., and Lau, V. K. N.: `Effect of carrier frequency offset on channel estimation for SISO/MIMO-OFDM systems', IEEE Trans. Wireless Commun., 2007, 6 (5), pp. 1854--1863 

\bibitem{Zeng2007}  
Zeng, Y., Leyman, A.R., and Ng, T. S.: `Joint semiblind frequency offset and channel estimation for multiuser MIMO-OFDM uplink', IEEE Trans. Commun., 2007, 55 (12), pp. 2270--2278

\bibitem{Chen2008}  
Chen, J., Wu, Y. C., Ma, S., and Ng, T. S.: `Joint CFO and channel estimation for multiuser MIMO-OFDM systems with optimal training sequences', IEEE Trans. Signal Processing, 2008, 56 (8), pp. 4008--4019

\bibitem{Yilmaz}
Yilmaz, A.: `Cooperative diversity in carrier frequency offset', IEEE
  Commun. Lett., 2007, 11, (4), pp. 307--309

\bibitem{Yadav}
Yadav, A., Tapio, V., Juntti, M., and Karjalainen, J.: `Timing and frequency
  offsets compensation in relay transmission for {3GPP} {LTE} uplink', Proc. IEEE ICC, Cape Town, South Africa, {M}ay 2010, pp. 1--6

\bibitem{Dabora08}
Dabora, R., and Servetto, S.: `On the role of estimate-and-forward with time
  sharing in cooperative communication', IEEE Trans. Inform. Theory, 2008, 54, (10), pp. 4409--4431

\bibitem{Ng2006}
Ng, C., Maric, I., Goldsmith, A., Shamai, S., and Yates, R.: `Iterative and
  one-shot conferencing in relay channels', Proc. IEEE Information Theory
  Workshop, {P}unta del {E}ste, {U}ruguay, March 2006, pp. 193--197

\bibitem{Draper03}
Draper, S. C., Frey, B. J., and Kschischang, F. R.: `Iterative decoding of a
  broadcast message', Proc. Allerton Conf. on Communication, Control,
  and Computing, {M}onticello, {USA}, 2003

\bibitem{Wittneben}
Kuhn, M., Wagner, J., and Wittneben, A.: `Cooperative processing for the {WLAN}
  uplink', Proc. IEEE WCNC, {L}as {V}egas, {USA}, {M}arch 2008, pp. 1294--1299

\bibitem{JiangGC10}
Jiang, J., Thompson, J. S., and Grant, P. M.: `Design and analysis of
  compress-and-forward cooperation in a virtual-{MIMO} detection system', Proc. 2010 IEEE GLOBECOM Workshop on Heterogeneous, Multi-hop, Wireless and
  Mobile Networks (HeterWMN), {D}ec. 2010, pp. 126--130

\bibitem{Kim}
Kim, T., Skoglund, M., and Caire, G.: `Quantifying the loss of compress-forward
  relaying without {W}yner-{Z}iv coding', IEEE Trans. Inform. Theory, 2009, 55, (4), pp. 1529--1533 

\bibitem{Gersho}
Gersho, A., and Gray, R. M.: `Vector quantization and signal
  compression', (Kluwer Academic Publisher, 1992)

\bibitem{Shanbehzadeh}
Shanbehzadeh, J., and Ogunbona, P.: `On the computational complexity of the {LBG}
  and {PNN} algorithms', IEEE Trans. Image Processing, 1997, 6, (4), pp. 614 --616 

\bibitem{Zhang}
Zhang, J., Zheng, Y., Xiao, C., and Letaief, K. B.: `Channel equalization and symbol detection for single-carrier {MIMO} systems in the presence of multiple carrier frequency offsets', IEEE Trans. Veh. Tech., 2010, 59, (4), pp. 2021--2030 

\bibitem{Wu}
Wu, Y. C., Chaudhari, Q., and Serpedin, E.: `Clock synchronization of wireless
  sensor networks', IEEE Signal Processing Magazine, 2011, 28, (1), pp. 124--138 

\bibitem{Noh}
Noh, K.-L., Chaudhari, Q., Serpedin, E., and Suter, B.: `Novel clock phase offset
  and skew estimation using two-way timing message exchanges for wireless
  sensor networks', IEEE Trans. Commun., 2007, 55, (4), pp. 766--777

\bibitem{Elson}
Elson, J., Girod, L., and Estrin, D.: `Fine-grained network time synchronization
  using reference broadcasts', Proc. 5th Operating Syst. Design and
  Implementation Symp., Boston, USA, {D}ec. 2002, pp. 147--163

\bibitem{Roehr}
Roehr, S., Gulden, P., and Vossiek, M.: `Method for high precision clock
  synchronization in wireless systems with application to radio navigation', Proc. IEEE Radio and Wireless Symposium, {J}an. 2007, pp. 551--554

\bibitem{LTE}
ETSI TS 125 104 V8.7.0, Technical Specification Universal Mobile
  Telecommunications System ({UMTS}); Base Station ({BS}) radio transmission
  and reception ({FDD}) (3GPP TS 25.104 version 8.7.0 Release 8) Std., July
  2009.

\bibitem{3GPP}
3GPP TS 36.211 V8.5.0, Technical Specification Group Radio Access Network;
  Evolved Universal Terrestrial Radio Access (E-UTRA); Physical Channels and
  Modulation (Release 8) Std., Dec. 2008.

\end{thebibliography}
\end{document}